\documentclass[
reprint,
superscriptaddress,
longbibliography,
amsmath,
amssymb,
aps,prb]{revtex4-2}
\usepackage{gensymb}
\usepackage{graphicx}
\usepackage{dcolumn}
\usepackage{bm}
\usepackage{xcolor}
\usepackage[mathlines]{lineno}
\usepackage[normalem]{ulem}
\usepackage[
			colorlinks = true,
            linkcolor = blue,
            urlcolor  = blue,
            citecolor = blue,
            anchorcolor = blue
            ]{hyperref}
\newcommand\rxout{\bgroup\markoverwith{\textcolor{red}{\rule[.5ex]{2pt}{.6pt}}}\ULon}

\begin{document}

\title{Ultra-broadband acoustic ventilation barriers via hybrid-functional metasurfaces}

\author{Ruizhi Dong}
 \thanks{These authors contributed equally to this work.}
\author{Dongxing Mao}%
 \thanks{These authors contributed equally to this work.}
\affiliation{Institute of Acoustics, Tongji University, Shanghai 200092, China}
\author{Xu Wang}
\email{xuwang@tongji.edu.cn }
\author{Yong Li}
\email{yongli@tongji.edu.cn}
\affiliation{Institute of Acoustics, Tongji University, Shanghai 200092, China}
\affiliation{College of Architecture and Urban Planning, Tongji University, Shanghai 200092, China
}
\date{\today}

\begin{abstract}
Ventilation barriers allowing simultaneous sound blocking and free airflow passage are of great challenge but necessary for particular scenarios calling for sound-proofing ventilation. Previous works based on local resonance or Fano-like interference serve a narrow working range around the resonant or destructive-interference frequency. Efforts made on broadband design show a limited bandwidth typically smaller than half an octave. Here, we theoretically design an ultra-broadband ventilation barrier via hybridizing dissipation and interference. Confirmed by experiments, the synergistic effect from our hybrid-functional metasurface significantly expand the scope of its working frequencies, leading to an effective blocking of more than $90 \%$ of incident energy in the range of $650-2000 \mathrm{Hz}$, while its structural thickness is only $53 \mathrm{mm}$ $(\sim \lambda / 10)$. Our design shows a great flexibility on customizing the broadband and is capable of handling sound coming from various directions, which has potential in air-permeable yet sound-proofing applications.
\end{abstract}

\maketitle

\section{\label{sec:level1}INTRODUCTION}

In acoustic engineering, there remains a significant challenge for simultaneous sound proofing and free flow passage. Acoustic barriers impede airflow transport, while the conventional ventilation barriers equipping winding airflow paths with absorptive linings bring about a large pressure drop and hence cannot maintain a free flow of air\cite{RN1,RN2,RN3}. Acoustic metasurface, as a research area that has recently generated a proliferation of work, have garnered much attention owing to their unique functional characteristics and vanishing size, and paved a new pathway for researchers to develop effective solutions for a wide variety of applications such as deep sub-wavelength focusing/imaging\cite{RN4,RN5}, one-way sound transportation\cite{RN6,RN7}, anomalous refraction and reﬂection\cite{RN8,RN9,RN10,RN11,RN12,RN13,RN14,RN15,RN16,RN17,RN18,RN19}, and compact absorbers\cite{RN20,RN21,RN22,RN23,RN24}. Ongoing advances in acoustic metasurface unlock possibilities for air-permeable barriers. These hollow-carved-board designs usually consist of periodically arranged hollow units to ensure sufficient air circulation. Based on local resonance\cite{RN25,RN26,RN27,RN28,RN29} (Helmholtz resonators, membranes, quarter-wavelength tubes, etc.) or Fano-like interference\cite{RN30,RN31,RN32}, they break the limitations of adjustable large-scale waves at subwavelength scales and enable a low frequency noise impeding in such an unprecedented open way. However, the underlying working mechanism implies that they serve narrow working frequency ranges around the resonant or destructive-interference frequencies. Efforts made to broaden the working frequency range either show a limited bandwidth\cite{RN29}, or a largely degraded airflow passage\cite{RN33}. It is shown recently that, by coupling multiple lossy resonators, a ventilation barrier can achieve broadband absorption at low frequencies\cite{RN34}. Besides, a barrier made from hollow-out helical meta-units was proposed, which features a consistently quasi-Fano-like interference in the mid-high frequency range and hence a broadband insulation\cite{RN35}. Although these designs\cite{RN34,RN35} show a remarkable progress on dealing with the broadband issue, the bandwidth is still limited, which are usually smaller than half an octave if evaluated by the range of blocking more than 90\% of incident energy.

Considering that noise usually covers a wide spectrum, designing an ultra-broadband ventilation barrier is still in pursuit. Here, via hybridizing dissipation and interference, we theoretically design and experimentally demonstrate an ultra-broadband ventilation barrier. The designed meta-unit is a hollow-out cylinder, which features a very similar profile to the previous work\cite{RN35} with almost the same outer and inner diameter, as well as the total thickness. Distinct from the so-far broadband designs\cite{RN34,RN35,RN36,RN37}, by harvesting the synergistic effect from the proposed hybrid-functional meta-unit, the scope of its working frequencies is significantly expanded. Such a barrier effectively blocks more than $90 \%$ of incident sound energy in the range of $628-1400 \mathrm{Hz}$ coming from various directions, while the structural thickness is only $53 \mathrm{mm}$ (approximately $\sim \lambda / 10$). By adding circumferential partitions to the meta-unit, we further achieve a expanded bandwidth (90\% sound energy insulation in the range of $650-2000 \mathrm{Hz}$). Our designs suggest an efficient approach towards noise control engineering in flowing-fluid-filled circumstances.

\section{RESULTS AND DISCUSSION}

\subsection{The ventilation barrier for broadband sound blocking}

\begin{figure*}
\includegraphics[width=15cm]{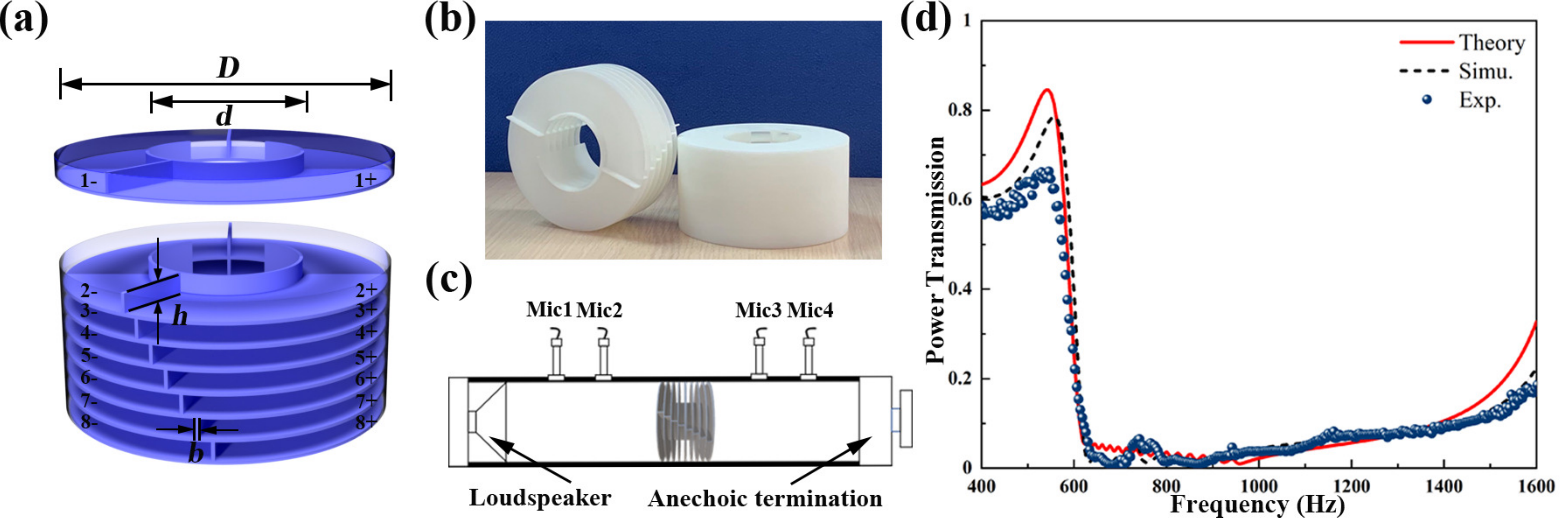}
\caption{\label{fig:wide}(a) The hollow-out design of the ultra-broadband ventilation barrier have eight layers. Each layer is composed of two resonators with different chamber depths, separated by a partition blade. These blades in the eight layers are gradually shifted, making the whole structure consisting of sixteen detuned resonators in total. All resonators are labeled, for example 1- indicates the left resonator in the 1st layer and 6+ refers to the right resonator of the 6th layer. For clarity, the 1st layer is detached to show the details of its inner structure. (b) A photo of the 3D-printed specimen of the designed meta-unit (right) and its inner structure (left). (c) Schematics of the experimental setup where the specimen is clamped in the impedance tube. (d) Theoretical (red line), simulated (black dashed line), and measured (blue dots) acoustic power transmission of the designed meta-unit with (\(D=100 \mathrm{mm}, d=44 \mathrm{mm}, h=5.5 \mathrm{mm}, b=1 \mathrm{mm}, H=53 \mathrm{mm}, \psi=142.5^{\circ}, \theta=5^{\circ} \text { and } S_{n}=63.4 \mathrm{mm}^{2}\)).}
\label{Figure 1}
\end{figure*}

Figure \ref{Figure 1}(a) shows the designed meta unit, which is a hollow-out cylinder stacked by eight functional layers. For clarity, the first layer is detached to show the details of its inner structure. Each layer is split into two resonant chambers, with the same opening areas but different chamber lengths that separated by a deflected blade. In order to harvesting the coherent coupling for the broadband issue\cite{RN24}, the openings of the resonant chambers are adjacent to each other (opening area of the resonant chambers \( S_{n}\)). The geometries of the eight layers are the same, except that the deflected angles are gradually shift (angle of 1st layer partition \(\psi\), angle shift \(\theta\)), resulting the whole structure is in fact composed of sixteen detuned resonant chambers. As it will be shown in following, each chamber will provide its own contribution (high-efficient dissipation or interferrence) and they together collectively provide a remarkable sound blocking over a wide frequency range, while the whole structure features a subwavelength thickness. The detailed geometries of the design meta-unit are illustrated in Figure \ref{Figure 1}(a), involving outer diameter $D$, inner diameter $d$, and the thickness of unit chamber $h$. The thickness the inner and outer cylindrical shells, the top and bottom plates, as well as the partition blades, is $b$. The overall thickness of the meta-unit is $H(H=8 h+9 b)$.

\subsection{\label{sec:citeref}The effective model of the ventilation barrier}
The acoustic performance of the proposed meta-unit can be characterized by its transfer matrix $\mathrm{T}_{0}$, which relates the state vectors of the sound fields at the input and output parts across the meta-unit, as $\{p, v\}_{i n}^{T}=\mathbf{T}_{0}\{p, v\}_{\text {out}}^{T}$. Notice that the overall performance of the meta-unit is determined by both the cross-sectional mutations at the input and output interfaces, as well as the sixteen side-branched resonant chambers. Moreover, although the resonant chambers are stacked layer by layer (the center high the resonant chamber n$_{+/-}$ shown in Fig.1(a) can be written as  $\left.z_{n}=b / 2+(n-1 / 2)(h+b), n=1,2, \ldots, 8\right)$, owing to that the meta-unit features a subwavelength thickness so that the interval of these chambers is in a deep-subwavelength scale, one can further assume that all these resonant chambers shares the same location at the half height of the meta-unit (i.e. $z_{n}=H / 2, n=1,2, \ldots, 8$). Therefore, the transfer matrix $\mathrm{T}_{0}$ can be written as $\mathrm{T}_{0}=\mathrm{T}_{f} \mathrm{T}_{a} \mathrm{T}_{r}$. Hypothetically, this is equivalent to a rigid-walled shrunken-cross-sectional tube (height \textit{H}), with all these resonant chambers now mounted at its bisecting plane (see the details in Supplementary Note). Here, the matrix $\mathrm{T}_{f / r}$ describes the contribution of cross-sectional mutation of the first/latter half the mutational-cross-sectional tube, which is given by  
\begin{equation}
\mathbf{T}_{f / r}=\left[\begin{array}{cc}
\cos \left(k_{0} L_{C}\right) & j \sin \left(k_{0} L_{c}\right) / \phi_{0} \\
j \phi_{0} \sin \left(k_{0} L_{C}\right) & \cos \left(k_{0} L_{C}\right)
\end{array}\right]
\label{eq1}
\end{equation}
where \textit{ k}$_{0}$ is the wavenumber of acoustic waves in air, \textit{ \(\phi\)}$_{0}$ denotes the ratio of the open area in the meta-unit ($\phi_{0}=d^{2} / D^{2}$), and $L_{c}=(0.5 \times H+\Delta H)$ is the effective length of the first/latter half the mutational-cross-sectional tube taking into account the end correction $\Delta H$. Here the end correction can be set as $\Delta H=0.425 d\left(1-1.25 \sqrt{\phi_{0}}\right)$ by considering the radiation impedance due to the change in cross section in the process of sound propagation\cite{RN22}. The matrix  represents the contribution from the sixteen side-branched resonant chambers. Such a closely arranged resonant chambers support a strong coherent coupling effect, leading the resonant aggregate treated as an overall coupled system whose acoustic impedance can be calculated as follows. First, one can deﬁne the acoustic impedance of each individual resonant chamber referring to the overall coupled system by replacing all other resonant chambers by hard walls. In this way, the acoustic impedance of the resonant chamber n$_{+/-}$ referring to the overall coupled system can be expressed as $Z_{\mathrm{n} \pm}=-j \rho_{c} c_{c} \cot k_{c} L_{\mathrm{n} \pm} /(\xi \rho c)$\cite{RN24}, where $k_{c}, \rho_{c}$ and $c_{c}$ are the wave number, air density, and sound speed in the resonant chambers, respectively. Considering the intrinsic loss induced by viscous and thermal boundary layers in these narrow chambers, these parameters $\left(k_{c}, \rho_{c}, c_{c}\right)$ are now complex. $L_{n \pm}$ are the effective length of the resonant chambers $n_{+}$ and $n_{-}$, $\xi$ is the ratio of the cross-sectional area of the resonant chamber to its opening area. Then, we can obtain the overall acoustic impedance of these sixteen resonant chambers, which reads 
\begin{equation}
Z_{a}=\left[\sum \frac{h}{2 H}\left(\frac{1}{Z_{n+}}+\frac{1}{Z_{n-}}\right)\right]^{-1} (n=1,2, \ldots, 8)
\label{eq2}
\end{equation}
Eq.(\ref{eq2}) gives the impedance of the coherently coupled sixteen resonant chambers. By further considering the cross-sectional area of the mutational-cross-sectional tube S and the total opening area of all resonant chambers $S_{a}\left(S_{a}=16 S_{n}\right)$, $\mathbf{T}_{a}$ can be given by    
\begin{equation}
\mathbf{T}_{a}=\left[\begin{array}{cc}
1 & 0 \\
S_{a} /\left(S Z_{a}\right) & 1
\end{array}\right]
\label{eq3}
\end{equation}
Based on Eqs. (\ref{eq1}) and (\ref{eq3}), the transfer matrix T$_{0}$ for the designed meta-unit is finally derived. Then, the sound transmission coeﬃcient $(\tau)$ of the meta-unit can be straightforwardly predicted by the formula 
\begin{equation}
\tau=\left(2 /\left|t_{11}+t_{12}+t_{21}+t_{22}\right|\right)
\end{equation}
where $t_{i j}(i, j=1,2)$ is the elements of the matrix $\mathrm{T}_{0}$.
\subsection{ Numerical simulations and Experimental demonstration of the ventilation barrier}%
The designed meta-unit is numerically simulated by using the pressure acoustic module of COMSOL Multiphysics, in which the frequency domain study is performed to calculate the sound transmission through the meta-unit. To estimate its low-frequency behavior under plane wave incidence, the designed meta unit is emplaced in a cylindrical waveguide, and both the waveguide and meta-units are considered acoustically rigid. Both ends of the waveguide are set as plane-wave radiation boundaries while the incident wave is only excited at one side. To consider the losses due to the viscous and thermal dissipation in these resonant chambers, the narrow region acoustics model is employed.

The experiment is carried out to verify the designed meta-unit.Figure \ref{Figure 1}(b) shows the sample of the meta-unit made of photosensitive resin via 3D printing. The transfer matrix method is adopted to measure the power transmission of the meta-unit by using Sinus type-1401 impedance tube, where the sample is ﬁxed ﬁrmly with clamps in the impedance tube. The experimental setup shown in Figure \ref{Figure 1}(c) is in consistent with that adopt in numerical modelling. 

Figure \ref{Figure 1}(d) shows the power transmission of the meta-unit. Good agreement among theoretical calculation, simulation and experiment demonstrates the eﬀectiveness of the our design. Such a ventilation barrier blocks more than $90 \%$ of incident sound energy i.e.$10 \mathrm{dB}$ in $T L$ (transmission loss), in the range of $628-1400 \mathrm{Hz}$ (experimental data). Notice that, the designed mete-unit has a very similar profile to the previous work\cite{RN34}. Both are hollow-out cylinder design with almost the same outer and inner diameter, as well as total thickness. However, the capability of the design meta-unit $T L>10 \mathrm{dB}$ in the range of $628-1400 \mathrm{Hz}$ shows a significant advantage over its precedent \((TL>10 \mathrm{ dB} \text { in the range approximately } 900-1400 \mathrm{ Hz})\).

\begin{figure*}
\includegraphics[width=15cm]{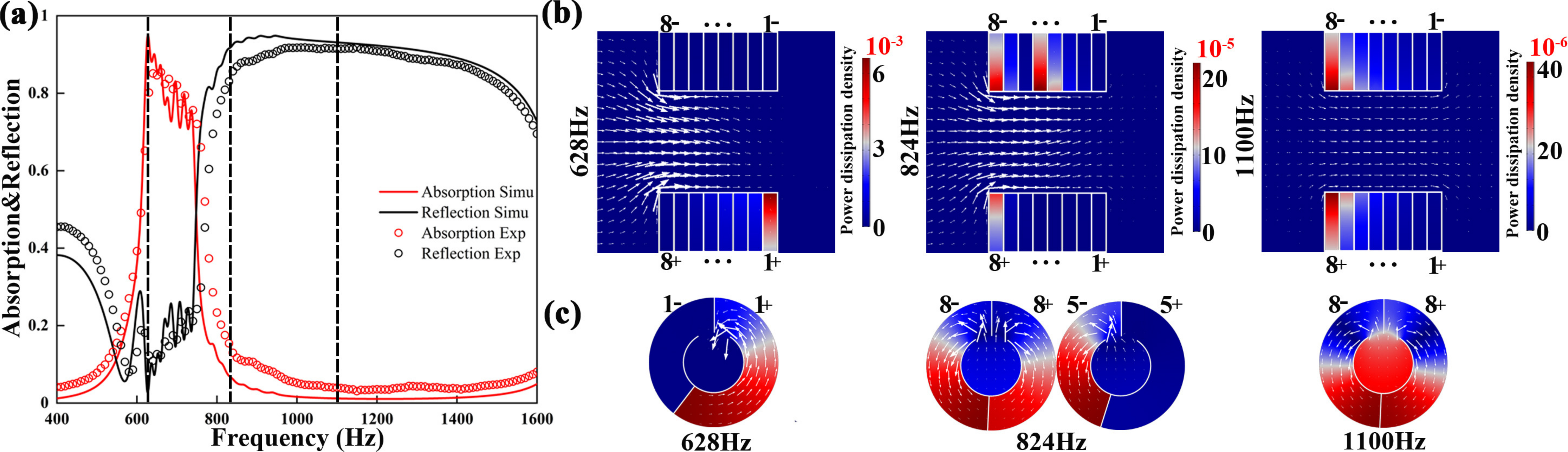}
\caption{\label{fig:wide}(a) The numerically calculated absorption (red line) and reflection (black line) coefficients of the meta-unit, demonstrating the such an ultra-broadband sound blocking is achieved via hybridizing sound absorption and interference. The black dashed lines refer to three frequency points, where the performance of the meta-unit is dominated by absorption ($628 \mathrm{Hz}$), absorption and interference ($824 \mathrm{Hz}$), and interference ($1100 \mathrm{Hz}$). (b) Simulated sound ﬁeld distributions at the three marked frequencies: $628 \mathrm{Hz}$ (top), $824 \mathrm{Hz}$ (middle), and $1100 \mathrm{Hz}$ (bottom). The left figures are the axial-plane view, where the color maps illustrate the power dissipation density. The color in red highlights those “active” resonators at these frequencies, while the max. value of the color bar clarifies the underlying working mechanism (absorption or interference). The white arrows indicate the sound intensity. Here sound impinges from left and the arrows diminish at the right, indicating strong sound blocking at these frequencies. (c) These figures are the radial-plane view of the specific layers where the “active” resonators belong to. Here the color maps illustrate the acoustic pressure distributions. The white arrows represent the local velocity streamlines, from which clear energy exchange between the resonators and the central orifice can be observed.}
\label{Figure 2}
\end{figure*}

\subsection{The working mechanism of the ventilation barrier}
The capability of the designed meta-unit on broadband sound blocking has been demonstrated. To further explore the underlying working mechanism, the absorption and reflection of the meta-unit have been numerically calculated and experimentally measured, as shown by Figure \ref{Figure 2}(a). This uncovers the fact that the such a wide working range is achieve via the synergy of absorption and reflection, i.e.a hybridized effect of dissipation and interference. Here we screen out three representative frequency points ($628 \mathrm{Hz}$, $824 \mathrm{Hz}$ and $1100 \mathrm{Hz}$), at which the performance of the meta-unit is dominated by absorption, absorption-reflection coaction, and reflection, respectively.
Figure \ref{Figure 2}(b) shows the numerical results of the sound field in the axial-plane view of the meta-unit at these representative frequencies, where the color maps illustrate the power dissipation density. The color in red highlights those “active” chambers, i.e. the chambers under strong resonance, at these frequencies. Notice that, the max. value of the color bar clarifies the underlying working mechanism. At 628Hz, the active unit (chamber $1_{+}$) provides the highest dissipation density $\left(\sim 10^{-3}\right)$, indicating a strong energy dissipation and a prominent absorption effect. At $824 \mathrm{Hz}$, the active units  (chamber $8_{+}$, and $8_{-}$) show a much lower dissipation density $\left(\sim 10^{-5}\right)$ so that the meta-unit works by the coaction of absorption and reflection. At $1100 \mathrm{Hz}$, the dissipation density of the active units $\left(\sim 10^{-6}\right)$ is three orders of magnitude lower than that at $628 \mathrm{Hz}$, which means that these active units support little sound absorption but strong reflection. The white arrows in Figure \ref{Figure 2}(b) represent the sound intensity, showing the energy flow distribution when sound passing through the meta-unit (from left to right). At $628 \mathrm{Hz}$, the obvious arrows at the input of the meta-unit gradually decays and finally disappears at the output, showing a strong energy dissipation through the meta-unit. Rather, at $1100 \mathrm{Hz}$, the diminished arrows at the input stem from the cancellation of quasi-equal strength of incident and reflected flux travelling in opposite directions. This is a clear evidence that now the meta-unit behaves as a strong reflector, while the contribution to absorption is negligible. When the meta-unit provide a hybridized function ($824 \mathrm{Hz}$), the distribution of sound intensity presents a pattern in between the former two extremes, as illustrated in the middle of Figure \ref{Figure 2}(c). Moreover, in these three cases, the arrows diminished at the output demonstrates the consistently strong sound blocking at these frequencies.

Figure \ref{Figure 2}(c) shows the cross-sectional view of the sound field in the layers those active working chambers located, at the three representative frequencies. Here the color maps and the arrows illustrate the sound pressure and local particle velocity distribution, respectively. Compared to those inactive ones (for example chamber $1_{-}$ at $824 \mathrm{Hz}$), the obvious arrows in the active chambers witness a massive energy exchange occurred between the chamber and the outer space, and the sound pressure distribution demonstrates the energy accumulation at the chamber terminal, both proofing a state of resonance. While these resonant chambers support whether high absorption or reflection depends on the matching degree of their acoustic resistance to air (a matching resistance to air suggesting a high-efficient dissipation and hence a strong absorption), which can be distinguished by their corresponding dissipation density shown in Figure \ref{Figure 2}(b).

Through the above analysis, it is demonstrated that the consistently sound blocking comes from the coaction of the sixteen coupled resonant chambers. Compared to the so-far broadband designs that utilize only interference\cite{RN34,RN36} or absorption\cite{RN35,RN37}, it is the synergistic effect makes our multi-functional design achieve sound blocking in such a broader frequency range. 
\begin{figure*}
\includegraphics[width=15cm]{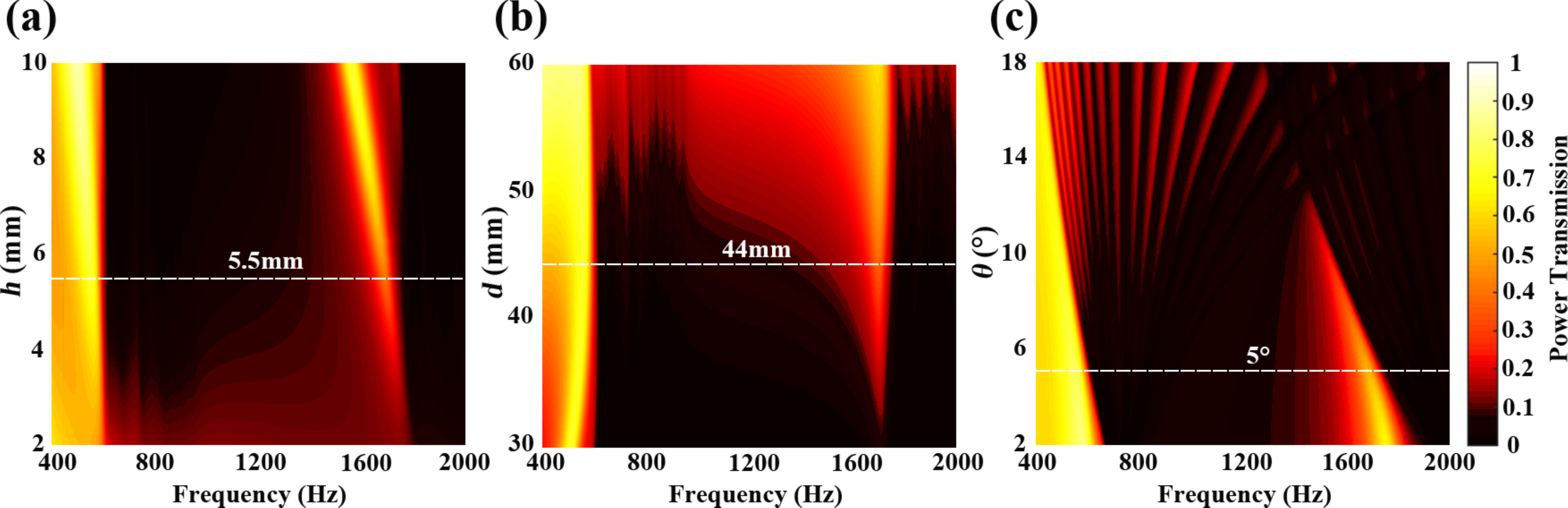}
\caption{\label{fig:wide}Power Transmission varying with (a) the thickness of each layer \(h\), (b) the inner diameter of the central oriﬁce \(d\), (c) the rotation angle of the partition blade in each layer cavity \(\theta\). The dark regions represent eﬀectively silenced zones. White dashed line in the colormap show the selected system parameters.}
\label{Figure 3}
\end{figure*}
\begin{figure*}
\includegraphics[width=15cm]{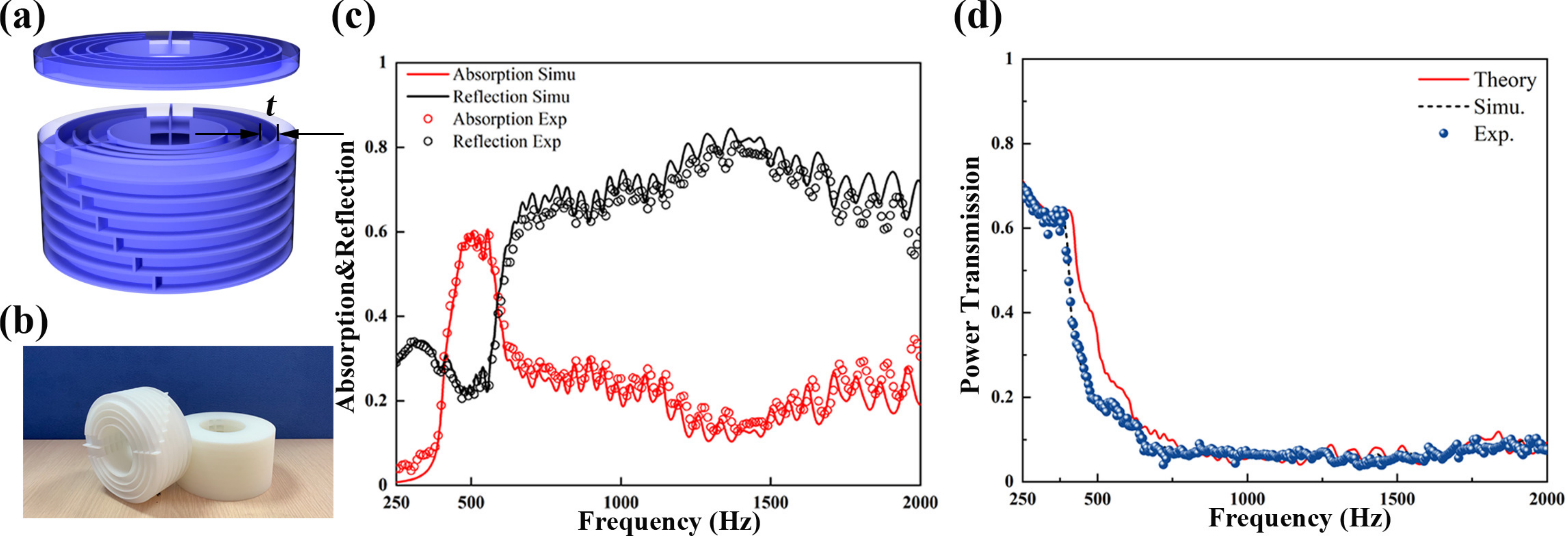}
\caption{\label{fig:wide}(a) The modified meta-unit featuring a extended working range. Compared to the designed shown in Figure \ref{Figure 1}(a), three additional circumferential partition blades are added in each layer, making the whole structure now is composed of 64 detuned resonators. (b) A photo of the 3D-printed specimen and its inner strucutre. (c) The numerically calculated absorption (red line) and reflection (black line) and measured absorption (red circle) and reflection (black circle) of the meta-unit. (d) Theoretical (red line), simulated (black dashed line), and measured (blue dots) Power transmission of the designed meta-unit(\(D=100 \mathrm{mm}, d=44 \mathrm{mm}, h=5.5 \mathrm{mm}, b=1 \mathrm{mm}, H=53 \mathrm{mm}, \psi=120^{\circ}, \theta=8^{\circ}, S_{n}=42.3 \mathrm{mm}^{2} \text { and } t=1 \mathrm{mm}\)).}
\label{Figure 4}
\end{figure*}

\subsection{ Numerical study of the ventilation barrier design strategies}%
The calculated power transmission is plotted as a 2D color map of the thickness \(h\) and the frequency, as shown in Figure \ref{Figure 3}(a). When other parameters remain unchanged, the bandwidth of the resonant cavity increases as the thickness increases. Until it is increased to $5.5 \mathrm{mm}$, the bandwidth reaches (power transmission less than 0.1) at $620-1400 \mathrm{Hz}$, after which the increase in thickness no longer plays a major role in the adjustment of the bandwidth. Next, the inner diameter of the hollow oriﬁce d is considered, which controls the unit’s open area ratio. Figure \ref{Figure 3}(b) shows the colormap with varying \(d\), as the inner diameter of the hollow oriﬁce \(d\) decreases the power transmission bandwidth gradually widens. The deflected angle of the partition in each layer is an arithmetic sequence, and the angle shift in each layer is \(\theta\). Figure \ref{Figure 3}(c) illustrates the case with varying  \(\theta\), by adjusting parameter  \(\theta\), it can achieve the expansion or compression of the sound insulation bandwidth. If the parameter  \(\theta\) was large enough, this unit can also realize the function of frequency selection.As aforementioned, the proposed ventilation barrier meta-unit provides great ﬂexibility on tailoring its acoustic property, leading to an adjustable frequency band by tuning the system parameters.

\subsection{  The modified ventilation barriers towards ultra-broadband sound blocking}%

In the above design, the deflected blade partitions a layer into two detuned resonant chambers. 
 \begin{figure}[b]
\includegraphics[width=5cm]{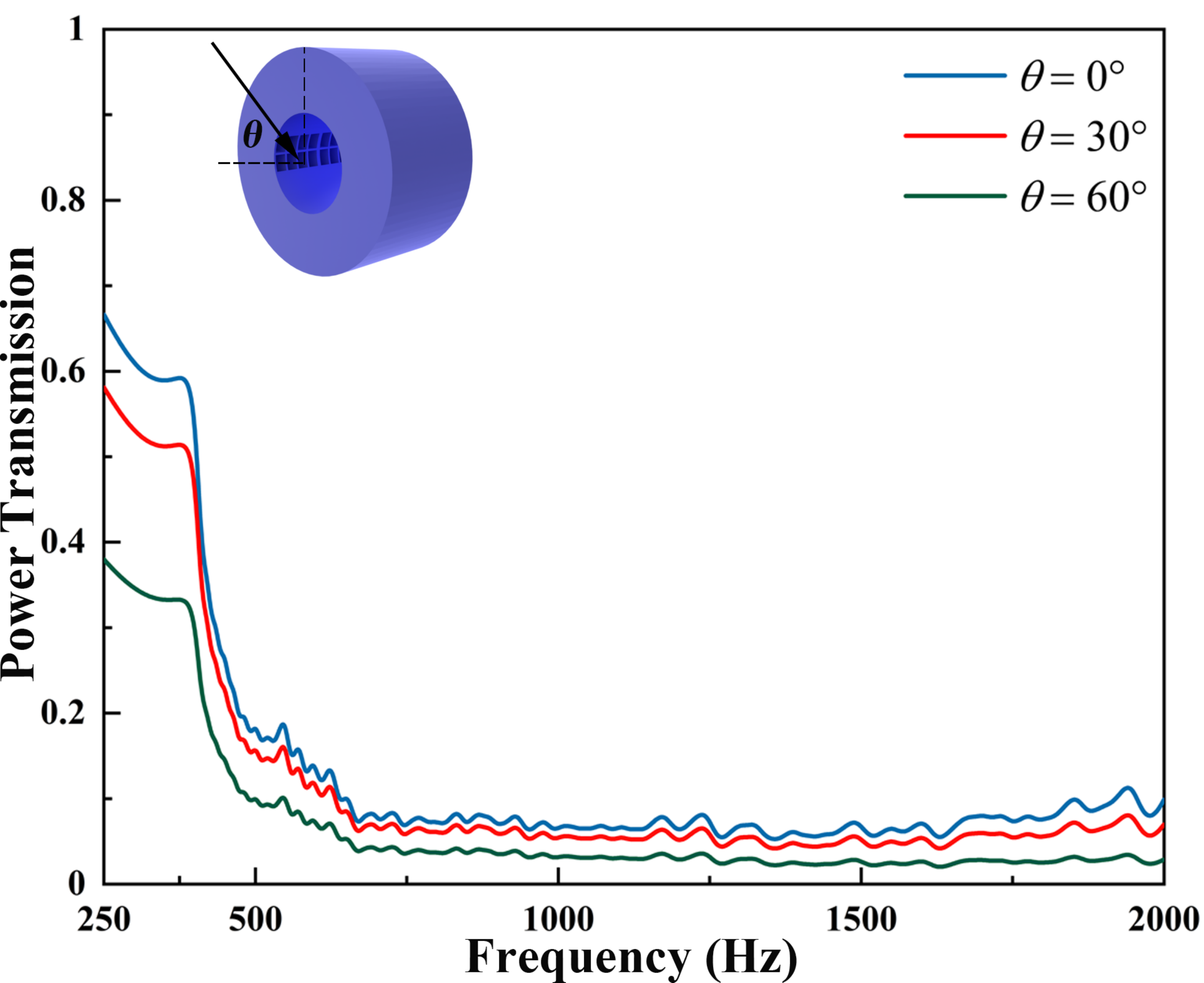}
\caption{\label{fig:epsart} Power transmission when plane waves come with incident angles of 0\degree(blue line), 30\degree (red line) and 60\degree(green line), respectively.}
\label{Figure 5}
\end{figure}
The deflected angle of these radial blades is gradually shifted from layer to layer, creating sixteen detuned resonant chambers that together collectively provide a broadband sound blocking. To further extend the working range of the meta-unit without changing its profile (outer diameter $D$, inner diameter $d$ and the overall thickness $H$), circumferential blades are deployed on each layer. As illustrated in Figure \ref{Figure 4}(a), now the modified meta-unit contains three circumferential blades evenly spaced (the partition spacing t) on each layer.Compared to the prototype shown in Figure \ref{Figure 1}(a), these circumferential partitions provide an additional degree of freedom, so that there are eight detuned chambers on each layers and sixty-four in total. Figure \ref{Figure 4}(b) shows the sample of the modified meta-unit made of photosensitive resin via 3D printing. This design, though features a more complicated inner structure, share the same underlying working mechanism as its prototype. The numerical simulation and experimental results shown in Figure \ref{Figure 4}(c) verifies that the modified meta-unit works via the synergy of dissipation and interference, where dissipation dominates at low frequencies (greater than $0.5$ at $450-580 \mathrm{Hz}$) while interference is prominent at higher frequencies. The theoretical, numerical and experimental data shown in Figure \ref{Figure 4}(d) witness the extended working range of the modified meta-unit, which blocking more than $90 \%$ of incident sound energy from $650 \mathrm{Hz}$ to $2000 \mathrm{Hz}$ (experimental data). Compared to its prototype ($628-1400 \mathrm{Hz}$), the additional circumferential partitions mainly contribute at higher frequencies. In practice, the functionality of an acoustic barrier should not restrict to only normal incidence. We numerically inspect the power transmission through the proposed acoustic barrier under oblique incidence (Figure \ref{Figure 5}). The simulation results show that the broadband characteristic is observable under a wide-angle range of incidence, while the sound blocking even becomes better under oblique incidence. This significantly expands the scope of the practicability of our design against various background noise fields.

\section{CONCLUSIONS}
In summary, we theoretically design and experimentally validate an ultra-broadband open barrier via a hybrid-functional acoustic metasurface. Different the previous broadband designs, the synergistic effect from our hybrid-functional metasurface significantly expand the scope of its working frequencies. Experiments are conducted to validate the proposed design, showing a consistently blocking of over $90 \%$ of incident energy in the range of $650-2000 \mathrm{Hz}$, while the structural thickness is only $53 \mathrm{mm}$ $(\sim \lambda / 10)$. The rich geometrical parameters oﬀer our design an adjustable frequency behavior. The broadband sound proofing in such an open yet compact manner pave the way for the solution in many particular scenarios calling for noise reduction and airflow passage simultaneously.

\begin{acknowledgments}
This work was supported by the National Science Foundation of China under Grant Nos.11774265, 11704255 and 11704284) and the Young Elite Scientists Sponsorship by CAST (Grant No. 2018QNRC001).
\end{acknowledgments}

\providecommand{\noopsort}[1]{}\providecommand{\singleletter}[1]{#1}%

\end{document}